# Small-world topology of functional connectivity in randomly connected dynamical systems

J. Hlinka,[1] D. Hartman,[1] and M. Paluš[1]
*Institute of Computer Science, Academy of Sciences of the Czech Republic, Pod vodarenskou vezi 2, 18207 Prague, Czech Republic*

(Dated: 19 June 2012)

Characterization of real-world complex systems increasingly involves the study of their topological structure using graph theory. Among global network properties, small-world property, consisting in existence of relatively short paths together with high clustering of the network, is one of the most discussed and studied. When dealing with coupled dynamical systems, links among units of the system are commonly quantified by a measure of pairwise statistical dependence of observed time series (functional connectivity). We argue that the functional connectivity approach leads to upwardly biased estimates of small-world characteristics (with respect to commonly used random graph models) due to partial transitivity of the accepted functional connectivity measures such as the correlation coefficient. In particular, this may lead to observation of small-world characteristics in connectivity graphs estimated from generic randomly connected dynamical systems. The ubiquity and robustness of the phenomenon is documented by an extensive parameter study of its manifestation in a multivariate linear autoregressive process, with discussion of the potential relevance for nonlinear processes and measures.

PACS numbers: 89.75.Fb,07.05.Kf,05.45.Tp,87.19.L-

In the field of complex systems study, new measurement and computational resources have lead to increased interest in analysis of large networks. These networks are observed across many disciplines spanning from social sciences through biology to climate research. For characterization of the structure of these networks, graph-theoretical measures have proven to be useful. These characteristics capture some global features of the network topology such as the density or level of clustering (tendency of neighbors of a node to be also neighbors to each other) as well as specific roles of important nodes serving as 'hubs' in the network. Among the interesting properties of many real-world networks belongs the small-world property, a global property of a network characterized by a relatively high level of clustering while conserving on average short paths among nodes of the network, compared to a random network of corresponding density. This small-world property has been related to some convenient properties of the network including efficiency of information transfer, and therefore reports of small-world architecture in real-world networks have received much attention. In some complex dynamical systems, including global climate or human brain, the knowledge of physical connections among its subsystems is far from perfect, and therefore other methods of characterizing interactions among these have been extensively applied. In particular, interactions among the areas are commonly quantified by a dependence measure such as the linear correlation between the local time series of variables of interest. This gives rise to the so-called functional connectivity matrix of a system. Applying the graph-theoretical approach to functional connectivity matrices has lead to reports of small-world properties of many real-world systems. However, as we document in this report, even for a simplistic dynamical system with linear dynamics and random coupling matrix the functional connectivity approach generates networks with small-world characteristics. These spurious detections of small-world topology is related to partial transitivity of functional connectivity measures such as the correlation coefficient.

## I. INTRODUCTION

Characterization of complex systems commonly includes the study of their structure using graph theory. This typically involves identification of the systems subunits (nodes of a network) and assessment of existence (or strength) of pair-wise relations among those, leading to representation of the system by a (weighted) graph. The local, mesoscale or global topology of the graph (or alterations thereof) are subsequently studied using various graph-theoretical measures with a goal of identifying systems properties key to its function. Among the most widely discussed properties are modularity, scale-freeness or small-world topology.

Since the paradigmatic publication of Watts and Strogatz in Nature[1], the small-world property entailing relatively short graph paths and high clustering has received much attention in many application areas dealing with complex systems. This includes such diverse fields as neuroscience[2] and climate science[3].

When working with such complex *dynamical* systems, for both practical and theoretical reasons, the links among nodes are commonly quantified by the dependence of the observed time series rather than the underlying physical or coupling network of connections. In the neuroscience community, this corresponds to the distinction between functional and structural (anatomical) connectivity. We adopt this useful terminology in this paper and suggest the general term *functional connectivity approach* to construction of graph representation of complex dynamical system by deriving the links through quantification (and potential thresholding) of the pair-wise statistical dependence measure (typically correlation) among the local time series of activity or other key variable of the nodes of the studied system.

As the exact values of quantitative graph-theoretical indices crucially depend on parameters such as network size or density, the interpretation of graph-theoretical properties of real-world networks is usually based on the comparison with 'corresponding' random graph models of Erdős–Rényi[4] or Maslov–Sneppen[5].

Some complications with the outlined analysis approach have recently been suggested in literature, particularly discussing the biasing effects related to sampling problems such as spatial oversampling[6] and finite size temporal samples and their autocorrelation[7].

However, we argue that there is a more fundamental problem with intepretation of the graph-theoretical properties of functional connectivity matrices. In particular, functional connectivity matrices are biased towards a specific structure due to their construction method; irrespective of particular time series length and/or sampling parameters. We focus particularly on functional connectivity matrices constructed using linear correlation as a measure of dependence. A specific example of an in-built bias of a linear correlation matrix is its 'weak' transitivity property: for any three random variables $X, Y, Z$ a strong positive correlation between two pairs of them implies a positive correlation within the third pair, including specific 'hard limitations', such as that $\rho_{XY}^2 + \rho_{YZ}^2 > 1$ implies $\rho_{XZ} > 0$, i.e. positivity of the third correlation coefficient (for a proof of a general form of this inequality see[8]). Such implicit dependence among the entries of the correlation matrix has been commonly overlooked in the interpretation of graph analysis of functional connectivity matrices.

In the presented study, we focus on the effects of this dependence on the graph-theoretical properties of the functional connectivity matrices. In particular, we show that there are tendencies towards specific graph structures in functional connectivity matrices computed from activity time-series of randomly structurally connected networks. A prime example of such an effect is the increased level of clustering, further affecting the estimates of small-world indices, potentially leading to serious misinterpretation of real-world data.

## II. METHODS

### A. Functional connectivity

Functional connectivity is most commonly quantified by the linear (Pearson's) correlation coefficient. While also non-linear connectivity measures are in use, for data with approximately Gaussian distribution is linear correlation practically sufficient[9,10]. The correlation matrix can be subsequently transformed into an unweighted graph ('binarized') by choosing a threshold and assigning links only to pairs of nodes with over-threshold correlation. In practice, the threshold is commonly chosen adaptively to achieve a predefined density of the resulting graph, although statistical testing may be applied as well[11–13].

### B. Graph-theoretical characteristics

Formally, in the graph-theoretical approach a network is represented by an unweighted graph $G = (V, E)$, where $V$ is the set of nodes of $G$, $N = \#V$ is the number of nodes and $E \subseteq \binom{V}{2}$ is the set of the edges of $G$. In some cases a separate function $w : E \to \mathbb{R}$ can be used to extend the graph concept by defining a weighted graph $G = (V, E, w)$, however throughout the current paper we only deal with unweighted graphs.

For any pair of nodes $i, j \in V$ we define an edge indicator $a_{i,j} = 1$ if and only if $\{i, j\} \in E$ (the graph contains an edge linking nodes $i$ and $j$), otherwise $a_{i,j} = 0$, node degree as $k_i = \sum_{j \in V} a_{i,j}$, and finally a path of length $\ell$ from node $i$ to node $j$ as a sequence of edges $\{m_0 = i, m_1\}, \{m_1, m_2\}, \ldots, \{m_{\ell-1}, m_\ell = j\}$ where all $m_i \in V$ are distinct nodes of the graph. The length of the shortest path between nodes $i$ and $j$ is denoted as $d_{i,j}$.

A graph $G$ can be characterized by its global properties, including the average path length

$$L = \frac{1}{N(N-1)} \sum_{i,j} d_{i,j} \qquad (1)$$

and the clustering coefficient

$$C = \frac{1}{N} \sum_{i \in V} c_i, \qquad (2)$$

where $c_i$ denotes the local clustering coefficient defined for $k_i \geq 2$ as

$$c_i = \frac{\sum_{j,\ell} a_{i,j} a_{j,\ell} a_{\ell,i}}{k_i(k_i - 1)} \qquad (3)$$

and for $k_i \in \{0, 1\}$ as $c_i = 0$.

Note that for disconnected graphs, some pairs of nodes are not connected by a path and the respective $d_{i,j}$ is not well defined (or set to infinity). In this case we compute the average path length by averaging only over pairs for which a finite path exists, see also the discussion section.

In their original paper, Watts and Strogatz[1] suggested the term "small-world" for networks that have a similar average path length, but an increased clustering coefficient compared to a corresponding random graph (which corresponds to a relative average path length and clustering coefficient $\lambda = \frac{L}{L_{rand}} \gtrsim 1$, $\gamma = \frac{C}{C_{rand}} \gg 1$) and also provided a simple generative model for such networks. The small-world property has been more recently proposed to be summarized in the small-world index $\sigma = \frac{\gamma}{\lambda} \gg 1$[14].

Although generally accepted, the interpretation of the small-world index has been critically discussed for various reasons including the potential confounding effects of spatial oversampling of the underlying system[6], finite-time estimation effects[7] or unclear interpretation of paths in the functional connectivity graphs[15]. In the following, we show that the functional connectivity approach with a common choice of dependence measure leads to a substantial upward bias in the small-world index $\sigma$. Understanding this phenomenon is key for accurate interpretation of experimental findings and their comparison across methodologies.

For simplicity, consider an autoregressive process of order 1 (AR(1)):

$$X_t = c + AX_{t-1} + e_t, \qquad (4)$$

where $c$ is a $N \times 1$ vector of constants, $A$ is a $N \times N$ matrix and $e_t$ is a $N \times 1$ vector of error terms. For simplicity we choose $c = \mathbf{0}_{N,1}$ and $e_t \sim \mathcal{N}(0,1)$ and $A = s(SC + \alpha \mathbb{I})/\lambda_{max}$, with the symmetric binary structural connectivity matrix $SC = SC(N,p)$ generated as a realization of the Erdős-Rényi model $G(N,p)$. In particular, each nondiagonal entry of the $N \times N$ SC matrix is assigned randomly and independently either with value 1 (edge exists, with probability $p$) or value 0 (edge does not exist, with probability $1-p$); diagonal elements are set to 0. $\mathbb{I}$ here denotes the identity matrix. The parameter $s \in (0,1)$ modulates the relative strength of the autoregressive and noise terms in (4) with the normalization by $\lambda_{max}$ – the largest (in absolute value) eigenvalue of the matrix $SC + \alpha \mathbb{I}$. The parameter $\alpha$ varies the relation between the autocorrelation and cross-correlation component of the AR process (4).

## C. Finite time series example

As a motivational example, we generate a finite sample of stochastic process with length $T = 300$ with parameter setting $p_{SC} = 0.1, N = 100, s = 0.1, \alpha = 2$, and compute the corresponding functional connectivity matrix $FC$ by binarizing the sample Pearson correlation matrix. The binarization threshold is chosen such that the density $p_{FC}$ of the binarized functional connectivity matrix is equal to the density $p_{SC}$ of the structural connectivity matrix (diagonal elements of the FC matrix are first set to zero). The respective matrices are shown in Figure 1.

A visual inspection shows that while the entries of the structural connectivity matrix are random and mutually independent, the functional connectivity matrix shows a specific structure. This can be quantified by the graph-theoretical measures. In this particular realization, we have $L_S = 2.157, L_F = 2.308, C_S = 0.1081, C_F = 0.2355$. As $SC$ and $FC$ have the same densities, $SC$ is effectively a realization of the Erdős-Rényi model corresponding to the density of $FC$, and we obtain the relative graph measures: $\lambda = 1.07, \gamma = 2.18, \sigma = 2.04$. The values indicate increased clustering and approximately conserved average path length with respect to a corresponding random graph. Together this signifies a small-world like structure of the functional connectivity matrix, even though the coupling structure of the generating system is completely random.

In view of the results of this simple experiment, it is important to ask what the significance of the findings of increased clustering and small-world structure in dependence matrices of real-world data is.

## D. Asymptotic behavior of correlation matrix: Parametric study settings

In the following we focus on investigating this effect in a more detail, studying its strength as a function of some of the most relevant parameters of the underlying process. For the sake of tractability we still limit ourselves to AR(1) process, but vary the following parameters: size of the network $N$, density $p_{SC}$ of the coupling matrix $SC$, balance of the autocorrelation and cross-correlation (by parameter $\alpha$), balance between the autoregressive and noise terms by parameter $s$ and the level of thresholding by changing the required density $p_{FC}$ of the matrix $FC$. While the strength of the effect is also dependent on the length of the sample $T$, the finite size of the sample is not crucial, as the upward bias does not vanish even asymptotically. The covariance matrix $\Sigma$ for the AR(1) process (4) is given by the infinite sum $\Sigma = \sum_{i=0}^{\infty} A^i A^{iT}$ (for derivation see e.g.[16]), which due to symmetry of the matrix $A = A^T$ leads to a Neumann series that converges to $\Sigma = (\mathbb{I} - A^2)^{-1}$. From this expression the correlation matrix is easily obtained by trivial normalizations using diagonal elements of $\Sigma$. Thus the choice of the rather trivial AR(1) model gives us the possibility to study asymptotic behavior of correlation matrices without additional effects of sampling or series length which are not negligible when the correlation matrices are estimated from time series.

Notwithstanding the dimensionality of the investigated parameter space, the parameter ranges were chosen to cover reasonably the parameter space, in particular we used values $N \in \{50, 200, 500\}$, $s \in$ $\{0.2, 0.5, 0.75, 0.9, 0.99\}$, $\alpha \in \{0, 1\}$. We further varied $p_{SC}$ and $p_{FC}$ logarithmically in 24 steps within the $(0, 1)$ interval – more exactly both variables are defined as $2^n$ where $n$ is an arithmetic progression from 0 to $-6.9$ with step $-0.3$. The lowest density was therefore smaller than 0.01.

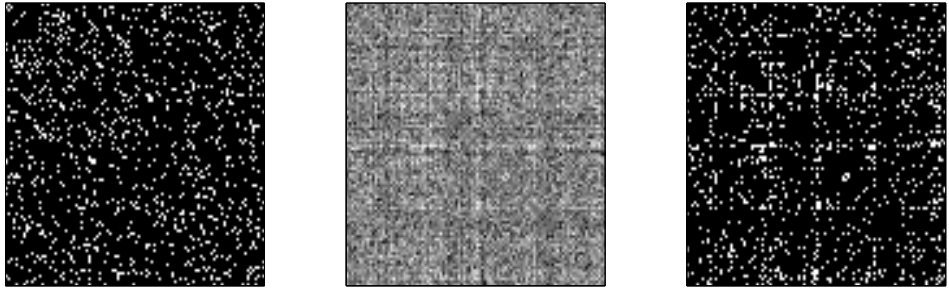

FIG. 1. An example of binary functional connectivity matrix (right) generated from random structural connectivity matrix (left) by thresholding the correlation matrix of AR-model generated time series (center, light shades of gray indicate higher correlation values). Network with $N = 100$ nodes shown. Note that the functional connectivity matrix shows a specific structure although the entries of the generating structural connectivity matrix were chosen randomly. See text for further details.

For robustness of evidence, for each parameter setting we compute 20 independent realizations of the coupling matrix, and each of the resulting matrices $FC$ is compared to its own corresponding realization of the random Erdős-Rényi matrix $G$ (and secondarily also to Maslov-Sneppen random graph model[5], see discussion). The computations were carried out using the NDW-Graph Toolbox (http://ndw.cs.cas.cz/software/ndw-graph), a C++/MATLAB toolbox for complex network analysis developed within the authors' NDW group and available to the public under GPL license.

### III. RESULTS

The simulations have shown that the studied effect (small-world property of functional connectivity matrix of a randomly connected dynamical system) is present throughout the covered parameter space, although with a variable strength. A representative selection of results is shown in Figure 2, obtained for network size $N = 500$ nodes, $s = 0.75$ and $\alpha = 1$. The results for other investigated parameter choices are qualitatively similar and are summarized for completeness in Figures 1 and 2 of the Supplemental Material[17].

Within the studied parameter range, we have observed values of small-world index $\sigma$ up to the order of hundreds. The effect has also proven statistically robust with respect to different realizations of the structural matrix $SC$. In particular, we observed relatively low spread of the $\sigma$ values, with $\sigma > 1$ in all 20 assessed realizations of the process (4) for overwhelming majority of parameter vector values. This corresponds to robust statistical significance in most cases (p-values $< 10^{-5}$, sign test of hypothesis of median equal to 1, no correction for multiple comparisons; similar results obtained for t-test). The only exceptions were observed for the case of exactly equal densities of structural and functional connectivity matrix, when this commong density was very low (e.g. only for $p_{FC} = p_{SC} \lesssim 0.03$ for the specific settings in Figure 2), where the $\sigma$ values were relatively close to 1; this special case is discussed later.

In general, $\sigma$ increases with increasing thresholding (that is decreasing density $p_{FC}$ of the $FC$ matrix). The dependence on the density of the underlying structural connectivity $p_{SC}$ is not monotonous (see Figure 2). Approximately, the effect is weakest if the density of the binarized functional connectivity matrix is the same as of the underlying structural connectivity matrix, but the depth of this minimum further depends on the value of parameter $\alpha$. The values of $\sigma$ are comparable for both $\alpha = 1$ and $\alpha = 0$, with the exception of the above-described non-monotonicity being more pronounced with $\alpha = 1$. With other parameters fixed, network size $N$ and strength of coupling $s$ do not affect $\sigma$ strongly within the sampled parameter region (see Supplemental Material[17] Figure 1).

However, for small networks (especially $N = 50$) and small densities of the $FC$ matrix $p_{FC} < 0.1$, the $\sigma$ is often not well defined, mostly due to zero values of the clustering coefficient of the corresponding random matrix. Moreover, in a part of the parameter space, the resulting graphs are not connected. This generally happens for small density, however the exact threshold is further modulated by network size and other network parameters.

The interpretation of some graph-theoretical measures for disconnected graphs and their comparison to connected graphs is not straightforward. In practice, for disconnected graphs the average path length is computed only for the largest component of the graph[1], or redefined as the mean of the finite elements of the distance matrix, omitting the unconnected node pairs. In the current study we adopt the latter approach, used previously in similar context[6].

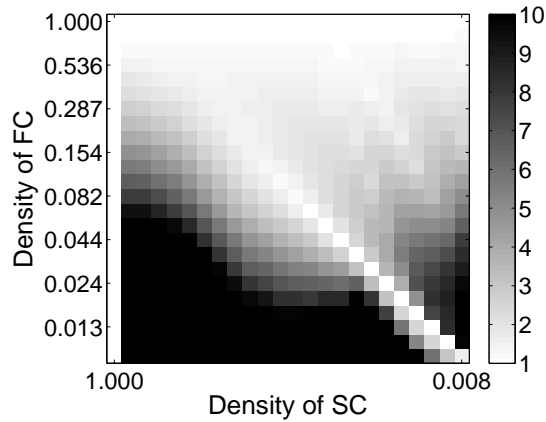 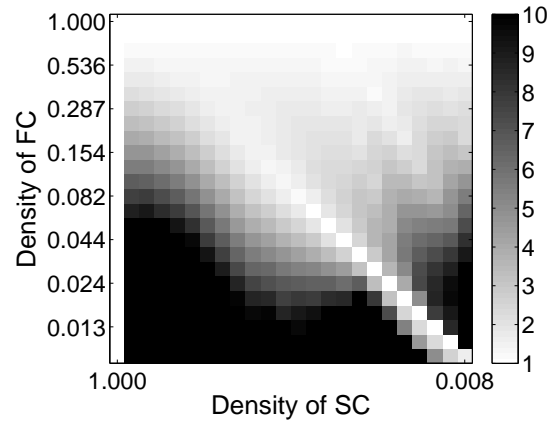

FIG. 2. Average values of the small-world index $\sigma$ of the binarized functional connectivity matrix $FC$ generated by AR model with random structural connectivity matrix $SC$. In most of the parameter space, $\sigma \gg 1$, suggesting small-world topology. Note marked dependence on values of the density of $SC$ and $FC$ (the latter modulated by adjusting the binarization threshold). Other parameter values: $s = 0.75$, $\alpha = 1$, $N = 500$. Logarithmic scale used both for the ordinate and the abscissa. Undefined values (for $SC = 1; FC \neq 1$) are shown as 0 (in white). (Results for other parameter settings are available in the Supplemental Material[17].)

FIG. 3. Average values of the relative clustering coefficient $\gamma$ of the binarized functional connectivity matrix $FC$ generated by AR model with random structural connectivity matrix $SC$. In most of the parameter space, $\gamma \gg 1$, contributing to small-world topology. Note the similarity to Figure 2, suggesting the critical role of increased clustering in the observed small-world topology of $FC$. Parameter values as in Figure 2. Logarithmic scale used both for the ordinate and the abscissa.

However, from theoretical perspective it could be argued that this could potentially have biasing effects on the small-world index estimates, supporting a conservative approach of only dealing with connected graphs. To check that our results are not driven only by loss of connectedness of the graphs with decreasing density or the related redefinition of average path, we plot separately the results obtained from connected graphs. While this severely limits the parameter space, even within this area $\sigma$ values attained values up to $\sim 10$ (see Supplemental Material[17], Figure 3).

As the admissible area of the parameter space monotonously grows with the network size, this documents that even the conservative connectedness requirement in principle does not limit the values of $\sigma$ and therefore the strength of the confounding effect.

The contributions of relative clustering coefficient $\gamma$ and relative average path length $\lambda$ to the observed small-world property is documented in Figures 3 and 4. The results show that the observed small-world property is mostly attributable to increased clustering with respect to corresponding random graphs (note the similarity between Figure 2 and Figure 3). In most of the studied parameter space, $\lambda$ is very close to 1 (i.e. $L$ similar to that of a random graph); lower values appear in general for parameter values that lead to disconnected graphs, typically with more components than random graph of corresponding density.

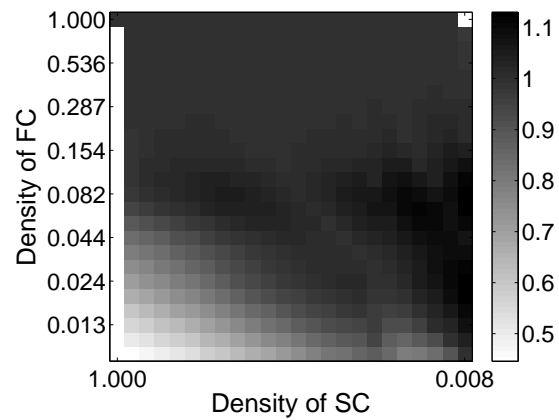

FIG. 4. Average values of the relative average path length $\lambda$ of the binarized functional connectivity matrix $FC$ generated by AR model with random structural connectivity matrix $SC$. In most of the parameter space, $\lambda \sim 1$, in agreement with small-world topology. Parameter values as in Figure 2. Logarithmic scale used both for the ordinate and the abscissa. See text for more discussion.

Because the small-world index $\sigma$ is computed by comparison with random networks, it is clear that the particular choice of the random network model may affect the $\sigma$ values. While Erdős-Rényi model is still standard for these applications, several recent studies used the Maslov-Sneppen model[5] that conserves not only the density but also degree distribution of the graph. We have explored the $\sigma$ values observed using the Maslov-

Sneppen random graph model using the same framework; the general observation was that of weakening, but not a complete mitigation of the effect (see Supplemental Material[17] Figure 4). This suggests that degree distribution in functional-connectivity-based graphs is indeed different from that of Erdős–Rényi model, yet it does not fully explain the specific properties of the functional connectivity matrix, in particular the increased level of clustering.

With other parameters fixed, we commonly observe the weakest effect if the resulting FC graph has the same density as the generating SC. This can be understood since in our model only the direct structural links are likely to generate the strongest (and quite homogeneous) temporal correlations. If the number of the functional connectivity links is set equal to that of structural links, the indirect and therefore weaker links are very likely to be discarded through the thresholding. The practical utilization of this fact is limited due to two points. Firstly, the density of underlying structural connectivity is typically not known in practice. Secondly, the strength of the structural links is commonly not homogeneous. Then, some indirect links can become stronger than some weaker direct links, causing increase in the minimum $\sigma$ value. This effect was observed in separate runs of the simulations that used Gaussian or uniform distribution for the strengths of structural links (see Supplemental Material[17], Figure 5 and 6).

## IV. DISCUSSION

We have shown that the approach of constructing network connectivity graphs from correlations of activity time series of local nodes leads to particular graph structures, characterized by increased clustering compared to common random graph models. This may lead to attribution of small-world properties to networks possessing purely random structural connectivity. By extensive mapping of strength of the effect as function of model parameters, we documented that this phenomenon is not a special theoretical case or a negligible effect, but is a rather pronounced and general phenomenon. While the coverage of parameter space was necessarily limited, it is sufficient not only to demonstrate the existence of the effect, but also its potential strength and existence of complex modulation by system parameters. The strength of the effect suggests it is of a scale relevant for interpretation of real-world results, considering that reports of small-world properties of real-world networks show $\sigma$ values comparable to those observed in our model – from the order of several units in smaller, relatively dense networks (e.g. $\sigma \sim 1.5$ in[18]) to several tens in larger and less dense networks (see e.g.[19]). Note that reports of small-world network topology based on functional connectivity analysis of measured time series is becoming increasingly used not only in brain imaging (see[20] and references therein), but also in other fields such as climate science[3,21] or economics[22].

As one of the most widely discussed concepts in the emerging field of complex network analysis, the small-world phenomenon has already received some attention regarding the methodology of its quantification. The two most notable attempts in this direction focused on the confounding effects of inadequate spatial sampling[6] and finite-size temporal window estimation of correlation coefficient[7]. Some other reasons for a careful interpretation of graph-theoretical analysis of functional connectivity matrices have been noted in recent reviews[15,23,24].

In the current study we bring a direct theoretical argument supported by extensive quantitative simulations proving that the interpretation problems regarding correlation-based small-world networks stem from the intrinsic properties of commonly used functional connectivity measures, rather than merely from inappropriate sampling. Even without the finite sample size problem, and with no spatial oversampling, functional connectivity matrices of randomly coupled trivial linear processes show small-world properties.

The current study utilizes linear statistical dependence measure (namely Pearson's correlation coefficient) and linear autoregressive model to study the consequences of the functional connectivity approach to construction of network graphs. A list of parameters of the model was varied, covering a large portion of the parameter space. Nevertheless, the computational requirements motivated some limitations of the parameter swipe. The increased clustering was documented across a multidimensional range of parameter values and for several choices of random distributions of structural links. While the effect was ubiquitous, its strength depended substantially on parameter values – in particular, the highest values were typically obtained for sparse networks, an observation in line with results reported in literature (see[25]). The accurate theoretical description of the parameter dependence of the strength of the effect for a reasonably general class of systems is a subject of future work, as it might help taking the effect properly into account in data-analysis.

Indeed, the AR(1) process analyzed here may serve only as a simplest first approximation for some real-world systems studied with the above discussed functional connectivity approach. We conjecture that since the origin of the phenomenon is in the partial transitivity property of correlation matrix of a dependent variable set, the details of the stochastic process and even the fact that the data are generated by a temporal stochastic process are not central to the phenomenon. In particular, the potential for spurious small-world observations may be also relevant to correlation-based networks constructed from other than time series data. Moreover, based on the elementary theoretical argument we conjecture that the main qualitative result generalizes to other dependence measures — to the extent to which these alternative measures hold a similar partial transitivity property as the linear correlation. Moreover, if the linear correlation is applied on nonlinear systems, the phenomenon should also appear — linear correlation may not be able to capture the dependences fully, but it will still show its specific properties. The detailed treatment of the related issues is a topic of further work.

Importantly, in the current paper, we do not challenge the specific findings of small-world structure of the *structural networks* underlying systems such as human brain. Rather, we only suggest that observations of small-world observations in *functional connectivity*, although potentially appealing, may not *per se* say much interesting about the underlying system, as these are probably ubiquitous in completely randomly connected dynamical systems, in a manner illustrated on a simplistic example of a linear dynamical system in this paper. More generally, this small-world bias is a specific example of the general notion that graph-theoretical properties of functional connectivity matrices should be interpreted differently than those of structural matrices.

Considering our results, in order to properly support the case that in a particular experimental dataset the observed small-world functional connectivity properties are significantly elevated, one would have to compare the experimental data with appropriate specific null-model (rather than that of Erdős–Rényi or Maslov–Sneppen) or provide a correction for the bias related to the functional connectivity approach.

Unfortunately, the null model of independent linear stochastic processes preserving length, frequency content, and amplitude distribution of the original time series that was proposed in Ref.[7], may not be suitable for testing hypothesis regarding network topology, since its alternative includes any set of dependent processes, irrespective of the topology of the pattern of their dependence. Rejection of the null hypothesis according to Bialonski et al.[7] gives statistical support for any dependence structure from the small-world topology on one side, to the randomly connected processes on the other side.

As clearly demonstrated in the simulations, the strength of the bias crucially depends on many parameters of the underlying system, including the density of links and the distribution of their values.

On the other side, researchers commonly turn to computing "at least" functional connectivity because they consider the full structure of the underlying dynamical system difficult (or impossible) to estimate from available data. Theferore the above named parameters of the underlying system, necessary for determining the size of the small-world bias, are not available in practical situations. This is why constructing a suitable test or correction procedure is not straightforward and constitutes on open problem for further research.

Of course, in many situations, the functional connectivity approach may still be suitable and remain in wide use; it is just that the interpretation should take properly into account the limitations of the approach.

## V. CONCLUSIONS

In conclusion, the increasingly used functional-connectivity approach to complex network graph construction bears an inherent strong bias towards small-world-like properties as measured by the conventional criteria. This should be taken into account when analyzing data using this approach in conjunction with graph-theoretical analysis in order to avoid erroneous interpretations. Ultimately, principled solutions should be sought to overcome this bias as well as other complications concerning the relation between structural and functional connectivity[26]. Dynamical models of complex networks are likely to play a key role in this challenge.

## ACKNOWLEDGMENTS

This study was supported by the Czech Science Foundation project No. P103/11/J068.